\documentclass{emulateapj}
\def\ergcm2s{\ifmmode {\rm\,erg\,cm^{-2}\,s^{-1}}\else
                ${\rm\,ergs\,cm^{-2}\,s^{-1}}$\fi}
\newcommand{\kms}{\ifmmode {\rm\,km\,s^{-1}}\else
                ${\rm\,km\,s^{-1}}$\fi}
\newcommand{\lya}{\ifmmode {\rm\,Ly\hbox{-}\alpha}\else
                Ly-$\alpha$\fi}
\newcommand{\msol}{M$_{\sun}$}
\newcommand{\oiii}{[O\,{\sc iii}]}

\begin{document}
\title{FIRST SPECTROSCOPIC MEASUREMENTS OF [O\,{\sc iii}] EMISSION FROM LYMAN - ALPHA SELECTED FIELD GALAXIES AT Z $\sim$ 3.1\altaffilmark{1}$^{,}$\altaffilmark{2}}
\author{Emily M. McLinden\altaffilmark{3}, Steven L. Finkelstein\altaffilmark{4}, James E. Rhoads\altaffilmark{3}, Sangeeta Malhotra\altaffilmark{3}, Pascale Hibon\altaffilmark{3}, Mark L. A. Richardson\altaffilmark{3}, Giovanni Cresci\altaffilmark{5}, Andreas Quirrenbach\altaffilmark{6}, Anna Pasquali\altaffilmark{7}, Fuyan Bian\altaffilmark{8}, Xiaohui Fan\altaffilmark{8}, Charles E. Woodward\altaffilmark{9}}
\altaffiltext{1}{The LBT is an international collaboration among institutions in the United States, Italy and Germany. LBT Corporation partners are: The University of Arizona on behalf of the Arizona university system; Istituto Nazionale di Astrofisica, Italy; LBT Beteiligungsgesellschaft, Germany, representing the Max-Planck Society, the Astrophysical Institute Potsdam, and Heidelberg University; The Ohio State University, and The Research Corporation, on behalf of The University of Notre Dame, University of Minnesota and University of Virginia.}
\altaffiltext{2}{Observations reported here were obtained at the MMT Observatory, a joint facility of the University of Arizona and the Smithsonian Institution.}
\altaffiltext{3}{School of Earth and Space Exploration,  Arizona  State University,  Tempe, AZ  85287}
\altaffiltext{4}{George P. and Cynthia Woods Mitchell Institute for Fundamental Physics and Astronomy, Department of Physics and Astronomy, Texas A\&M University, 4242 TAMU, College Station, TX 77843}
\altaffiltext{5}{INAF - Osservatorio Astrofisico di Arcetri, Largo E. Fermi 5, I-50125 Firenze, Italy; Max Planck Institut f\"{u}r extraterrestrische Physik, Postfach 1312, Garching 85741, Germany}
\altaffiltext{6}{ZAH, Landessternwarte, Universit\"{a}t Heidelberg, Königstuhl 12, D-69117 Heidelberg,Germany}
\altaffiltext{7}{Max-Planck-Institut f\"{u}r Astronomie, K\"{o}nigstuhl 17, D-69117 Heidelberg, Germany}
\altaffiltext{8}{Steward Observatory, The University of Arizona, 933 N Cherry Ave., Tucson, AZ 85721}
\altaffiltext{9}{Department of Astronomy, 351 Tate Laboratory of Physics, 116 Church Street, S. E., University of Minnesota, Minneapolis, MN 55455, USA}
\begin{abstract}
We present the first spectroscopic measurements of the [O\,{\sc iii}] 5007\AA\ line in two z $\sim$ 3.1 Lyman-alpha emitting galaxies (LAEs) using the new near-infrared instrument LUCIFER1 on the 8.4m Large Binocular Telescope (LBT).  We also describe the optical imaging and spectroscopic observations used to identify these \lya\ emitting galaxies.  Using the [O\,{\sc iii}] line we have measured accurate systemic redshifts for these two galaxies, and discovered a velocity offset between the [O\,{\sc iii}] and Ly-$\alpha$ lines in both, with the \lya\ line peaking 342 and 125\kms\ redward of the systemic velocity. These velocity offsets imply that there are powerful outflows in high-redshift LAEs.    They also ease the transmission of \lya\ photons through the interstellar medium and intergalactic medium around the galaxies.  By measuring these offsets directly, we can refine both \lya-based tests for reionization, and \lya\  luminosity function measurements where the \lya\ forest affects the blue wing of the line.    Our work also provides the first direct constraints on the strength of the [O\,{\sc iii}] line in high-redshift LAEs.  We find [O\,{\sc iii}]  fluxes of 7 and 36 $\times 10^{-17}$ erg s$^{-1}$ cm$^{-2}$ in two z $\sim$ 3.1 LAEs.
These lines are strong enough to dominate broad-band flux measurements that include the line (in this
case, K$_s$ band photometry).  Spectral energy distribution fits that do not account for the lines would therefore overestimate the 4000\AA\ (and/or Balmer) break strength in such galaxies, and hence also the ages and stellar masses of such high-z galaxies.
\end{abstract}


\section{INTRODUCTION}

The Lyman-$\alpha$ emission line is a highly efficient tool for
identifying and studying star forming galaxies at high redshifts.
This line can carry up to $ 6\%$ of the bolometric luminosity of a young
stellar population \citep{pp67} and is conveniently placed for observations by
ground-based optical observatories for $2 \la z \la 7$.  However, the
transmission of \lya\ emission is complicated by its resonant scattering
interaction with neutral hydrogen, both within the galaxy emitting the
line and in the surrounding intergalactic medium (IGM).  The \lya\
line is observed in about 25\% of z $\sim$ 3 -- 5 Lyman-break galaxies (LBGs)
\citep[e.g.,][]{stei00,dh07,rho09}, a percentage that may increase with increasing redshift \citep{shima06,star10}. This trend is supported by work at redshifts less than two, where the fraction of galaxies exhibiting \lya\ emission decreases at these lower redshifts \citep{re08,ha10,co10}. The 
\lya\ line is observed to have a characteristically asymmetric profile, with a sharp
cutoff on the blue side and a more extended wing on the red side
\citep[e.g.,][]{rho03}.  In Lyman-break selected galaxies, the peak
of the \lya\ line is typically redshifted by several hundred \kms\
with respect to interstellar absorption lines and/or nebular emission lines \citep{shap03,stei10}, whereas this measurement has not been made in \lya\ selected galaxies until this paper.

Besides being a useful tool for studying galaxy properties, \lya\
galaxies also offer unique and powerful probes of cosmological
reionization \citep[e.g.,][]{rho01,mr04,mr06,kash06,mcc09,day10}  The detailed
interpretation of these tests can be substantially affected by
velocity offsets between \lya\ and the systemic redshift, because a
redshifted line is less affected by the damping wing of \lya\
absorption from the IGM \citep{san04,mr06,dw10}.  

It is not sufficient to assume that the
velocity offsets seen in LBG samples hold for \lya\ selected
samples.  LAEs are typically less massive than presently available Lyman-break selected samples at similar redshifts \citep{ven05,gaw06,f07,pir07,nil}.  They should have correspondingly lower escape speeds, provided that stellar mass correlates broadly with dark matter halo mass.  Such a trend is discussed by \citet{gaw07}, where their sample of z $\sim$ 3.1 LAEs have typical stellar masses of 1 $\times$ 10$^{9}$ \msol\ and median halo masses of 7.9 $\times$ 10$^(10)$ \msol.  Gawiser et al. point out that these values are significantly smaller than those values for LBGs at z $\sim$ 3.1, which have stellar masses of $\sim$ 2 $\times$ 10$^{10}$ \msol\ and halo masses of $\sim$ 3 $\times$ 10$^{11}$ \msol\ \citep{shap, ad05}.  Galactic winds (and indeed many other astrophysical outflows) typically have flow speeds near the escape speed for the object, and the observed velocity offset of a \lya\ line is roughly double the wind speed \citep{v06}.  Additionally, the velocity offsets in Lyman-break samples are inversely
correlated with the \lya\ emission strength, as characterized by
equivalent width \citep{shap}, and the equivalent widths of
the \lya\ selected samples are much larger on average than those of
LBG samples.  Finally, \lya\ selected galaxies are typically
small in physical size \citep{bon09,bon10,ma10}.  

We present here the first direct measurements of the velocity offset
between \lya\ and nebular emission lines for typical \lya\ selected
galaxies.  Our measurements are based on a combination of
near-infrared spectroscopy with the new LUCIFER instrument on the Large
Binocular Telescope, and optical spectroscopy using Hectospec on the
MMT.  We selected targets for the study from a large area narrowband
survey conducted with the 90Prime camera on the 2.3m Bok telescope of
the Steward Observatory.

In section~\ref{sec:obs} we describe our observations and data
analysis methods.  We present our observational results in
section~\ref{sec:results}, and discuss their implications in
section~\ref{sec:discuss}. Where relevant, we adopt H$_{0} =$ 70 km s$^{-1}$ Mpc$^{-1}$, $\Omega_{m} = $ 0.3, $\Omega_{\Lambda} =$ 0.7 \citep{sper}.  Also we use the following vacuum wavelengths, 1215.67  \AA\ for \lya, 3729.875 \AA\ for [O\,{\sc ii}], 4862.683 \AA\ for H$\beta$  and 4960.295/5008.240 for [O\,{\sc iii}] from the Atomic Line List v2.04\footnote{http://www.pa.uky.edu/$\sim$peter/atomic/index.html}.  All magnitudes quoted are AB magnitudes unless otherwise specified.

\section{OBSERVATIONS AND DATA PROCESSING}
\label{sec:obs}
\subsection{Narrowband Survey - Observations and Data Reduction}
We completed a deep narrowband survey for LAEs at z $\sim$ 3.1 using the 90Prime Camera on the 2.3m Bok telescope at the Steward Observatory \citep{will04}.  The survey was completed in the COSMOS field centered at RA 10:00:28.6 and DEC $+$02:12:21.0 (J2000) \citep{cap07}.  The KPNO MOSAIC[O\,{\sc iii}] filter, centered at 5025 \AA, with a bandwidth of 55 \AA, was used to select Lyman alpha emission at redshifts z $\sim$ 3.1.  The data was obtained through time allocated by Steward Observatory in February 2007 (PI Finkelstein) and February and March 2009 (PI McLinden).  We have created a 1.96 deg$^2$ image, representing a total integration time of 16.67 hours.  
The complete details of this survey and the data reduction process will be highlighted in a forthcoming paper.

\subsection{Broadband Data}
We obtained publicly available broadband imaging data in CFHT u$^{\ast}$ and SDSS g$^{+}$ bands from the NASA/IPAC Infrared Science Archive\footnote{http://irsa.ipac.caltech.edu/data/COSMOS/datasets.html} to complement our narrowband survey.  The g$^{+}$ imaging data (v2.0) comes from Suprime-Cam \citep{mi02} on the Subaru 8.3m telescope.  The u$^{\ast}$ band imaging data (v5.0) comes from the MegaPrime/MegaCam\footnote{Based on observations obtained with MegaPrime/MegaCam, a joint project of CFHT and CEA/DAPNIA, at the Canada-France-Hawaii Telescope (CFHT) which is operated by the National Research Council (NRC) of Canada, the Institut National des Science de l'Univers of the Centre National de la Recherche Scientifique (CNRS) of France, and the University of Hawaii.} on the Canada-France-Hawaii 3.6m Telescope.  The 5$\sigma$ depth in a 3$\arcsec$ aperture in each band is 26.4 and 27.0  for the u$^{\ast}$ and g$^{+}$ bands, respectively \citep{cap07}. These broadband images were registered to our narrowband image using the IRAF tasks WCSMAP and GEOTRAN, which resamples the broadband images to match the coordinate system of the narrowband image (0.45 $\arcsec$/pixel).

\subsection{Candidate Selection from Narrowband and Broadband Data}
We used the SExtractor package \citep{ba96} to perform source detection in the narrowband and broadband images.   SExtractor was run in dual-image mode, first with narrowband image as both the detection and measurement image, and a second time with the narrowband as the detection image and broadband image as the measurement image.  
We selected LAE candidates based on the strength of their narrowband versus broadband excess as well as their colors as outlined in Rhoads $\&$ Malhotra (2001).

Namely, LAE candidates must be detected in the narrowband at the 6$\sigma$ level, their flux in the narrowband must exceed that in the broadband (g$^{+}$ band) by at least a factor of 2 and their narrowband flux must exceed their g$^{+}$ band flux at the 4$\sigma$ level. Finally, candidates must have flux in the filter bluward of of the \lya\ line (u$^{\ast}$ band) consistent with expected \lya\ forest absorption blueward of the \lya\ line and consistent with a u$^{\ast}$ - g$^{+}$ color $\ge$ 2.  Selection criteria are shown below in equations 1 -- 4.   
\begin{eqnarray}
f_{nb} / \delta f_{nb} & \ge& 6\\
f_{nb} / f_{g} & \ge& 2\\
f_{nb} - f_{g} & \ge&4\left(\delta f_{nb}^2 + \delta f_{g}^2\right)^{1/2}\\
f_{u} & \le& 10^{-4/5}f_{g} + \left(3\times \delta f_{u}\right)
\end{eqnarray}
Here f$_{nb}$ is the narrowband flux, $f_{g}$ is the g band flux, $f_{u}$ is the u band flux, $\delta f_{nb}$ is the flux error in the narrowband, $\delta f_{g}$ is the flux error in the g band, and $\delta f_{u}$ is the flux error in the u band. The second criterion requires that objects have \lya\ equivalent  widths $\ge$ 57.5 \AA.  The decision to require a 6$\sigma$ detection in the narrowband and to require a 3$\sigma$ non-detection blueward of the Lyman-break indicates that these are stringent selection criteria, meant to exclude false detections and low redshift interlopers.  We used isophotal magnitudes (MAG\_ISO) from SExtractor to measure the magnitudes and fluxes of each object.  Isophotal magnitudes were chosen because they have been found to produce the most accurate colors when SExtractor is run in dual-image mode \citep{ho05}.  Isophotal magnitudes are not measured within a fixed aperture for each object, but rather determines the magnitude from the number of counts in pixels above the user defined threshold and hence each object has a unique `aperture' in which its flux is measured.  For instance, the two LAEs with detected [O\,{\sc iii}] emission have extracted narrowband isophotal areas of 20.86 arcsecond$^2$ and 8.71 arcsecond$^2$ (later referred to as objects LAE40844 and LAE27878, respectively).   Similar selection criteria used at z=4.5 have typically yielded a spectroscopic success rate of 80\% (Dawson et al. 2004, 2007, Rhoads et al. 2003, 2005, Wang et al. 2009).

\subsection{Optical Spectroscopy - Observations and Data Reduction}
We obtained optical spectra of LAE candidates in January, February and April 2009, using the Hectospec multi-fiber spectrograph at the 6.5m MMT Observatory (a joint facility of the Smithsonian Astrophysical Observatory and the University of Arizona). Hectospec has a 1 deg$^2$ field of view and spectral coverage from 3650 - 9200 \AA.  The resolution of the instrument is $\sim$ 6 \AA.  Optical spectra are crucial for confirming that candidates are in fact LAEs at the correct redshift and not lower redshift interlopers and for determining the exact wavelength for the \lya\ line.

We reduced the Hectospec data and extracted 1D spectra using the External SPECROAD\footnote{http://iparrizar.mnstate.edu/$\sim$juan/research/ESPECROAD/index.php} pipeline developed by Juan Cabanela.  The External SPECROAD pipeline applies bias, dark and flat field corrections as well as wavelength calibration using He-Ne-Ar arc lamps. Typical residuals from the wavelength calibration are 0.15 \AA.

The optical spectra of the three objects chosen for near-infrared followup are shown in Figure 1.  These optical spectra confirm that these objects are in fact LAEs.  The spectra show strong \lya\ emission at the expected wavelength and the line displays the characteristic asymmetry expected for this line when emitted from a high-z source \citep{rho03,daw04,kash06}.  See section 3.1 for further discussion of this asymmetry.  Finally the spectra were checked for the presence of any other optical lines.  No additional emission lines were observed at the wavelengths where they might be expected for foreground [O\,{\sc ii}]  or [O\,{\sc iii}]  emission line objects.

\begin{figure}
\epsscale{0.4}
\plotone{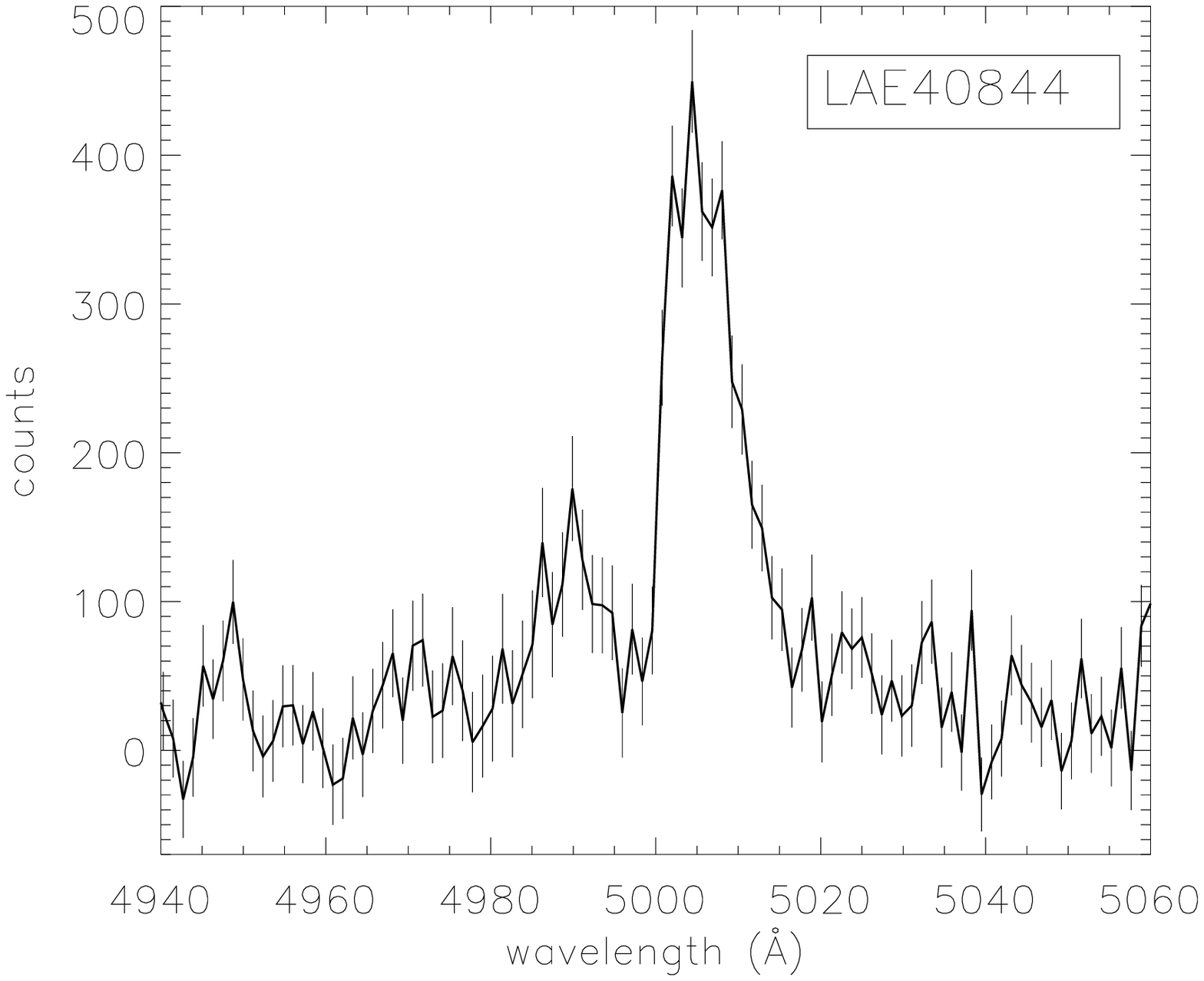}
\plotone{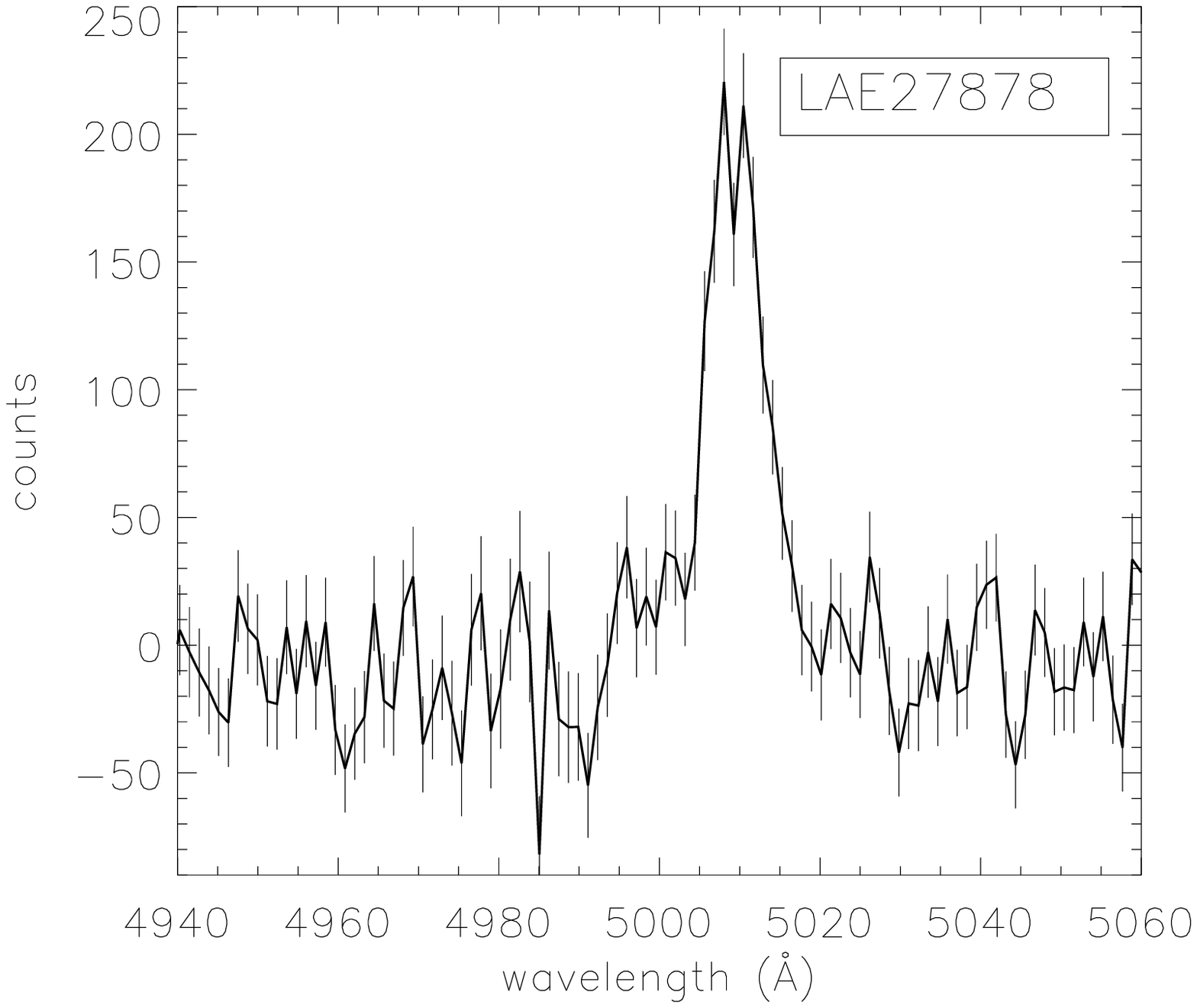}
\plotone{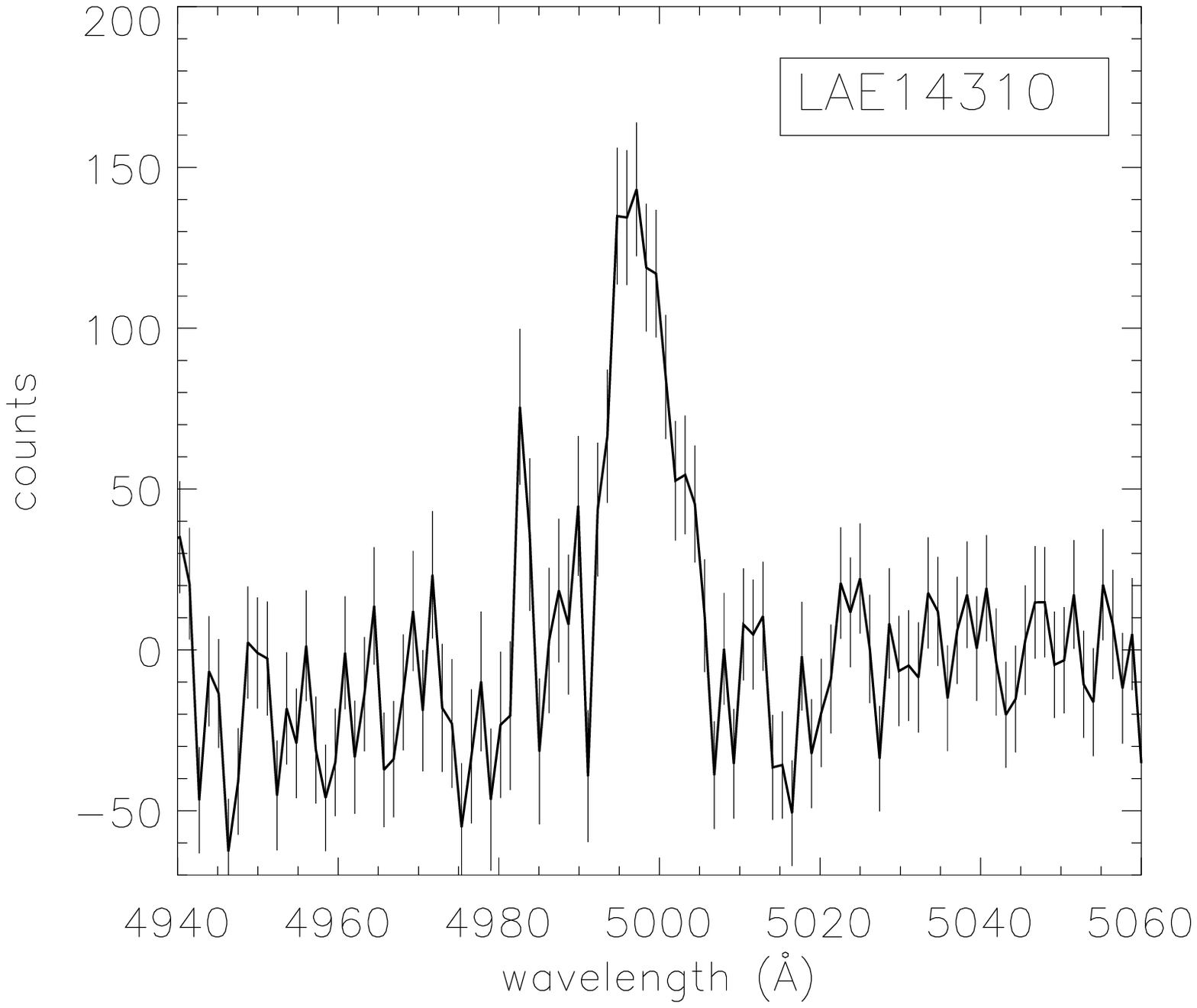}
\caption{Hectospec optical spectra - used for confirmation of the objects as \lya\ emitting galaxies at z $\sim$ 3.1.  LAE40844 has an additional feature, a `blue bump', at $\lambda \sim$ 4990 \AA, which is discussed in section 3.4. The spike seen near 4982 \AA\ is noise and not an additional feature.}
\end{figure}

\subsection{Near-Infrared Spectroscopy - Observations and Data Reduction}
Three of our brightest confirmed LAEs after Hectospec observations were observed in the near-infrared (NIR) using the new near-infrared instrument LUCIFER (LBT NIR Spectrograph Utility with Camera and Integral-Field Unit for Extragalactic Research) on the 8.4m LBT \citep{sei02,ag10}  The \lya\ line flux of these objects chosen for NIR followup, derived from their narrowband and broadband magnitudes, ranges from 0.94 -- 3.6$\times10^{-16}$ erg s$^{-1}$ cm$^{-2}$.   LUCIFER1 is the first of two planned NIR instruments  for the two 8.4m mirrors of the LBT.  LUCIFER1 currently operates on one mirror of the LBT and is capable of spectroscopy and imaging in the wavelength range 0.85$\mu$m -- 2.5$\mu$m. Our observations were performed in queue mode during LUCIFER's Science Demonstration Time in December 2009 and continued during the Arizona Queue in January and February 2010.  

We used the longslit mode of LUCIFER with a 1$\arcsec$ slit utilizing the H+K grating with 200 lines/mm and the N1.8 camera.   The image scale of the N1.8 camera is 0.25$\arcsec$/pixel.   We obtained 10 two-minute integrations for our brightest LAE (henceforth LAE40844).  Our second object (henceforth LAE27878) was observed over 20 four-minute integrations.  Our final object (henceforth LAE14310) was observed over 25 four-minute integrations.  The central wavelength in this setup is 1.93 $\micron$, and the spectral coverage spans essentially the full H and K band windows.  The spectral resolving power with the 4 pixel slit ranges from 940 near $1.6\micron$ to  1286 near $2.2 \micron$ or a resolution of $\sim$ 4.3 \AA/pixel.

We utilized the DOSLIT routine in IRAF \citep{val93} to reduce the 2D spectra.  To simplify reduction, our observations were designed so that a bright (R $\sim$ 12 -- 18) continuum source also shared the slit with each LAE.  This allows for a trace to be created using the bright continuum object.  The trace was then shifted along the spatial axis to extract the much fainter LAE, whose continuum emission is undetectably faint in individual exposures.  We performed flat fielding and dark correction before aperture extraction.  An aperture of 6 pixels was used for extraction.  Wavelength calibration, also performed as part of the DOSLIT task, was done using an argon lamp spectrum observed in the same setup as our science observations. After reduction, we averaged individual frames using the IRAF task SCOMBINE to produce a single averaged spectrum for each object.  Average RMS uncertainties from wavelength calibration for the two spectra with  [O\,{\sc iii}] detection were 0.64 \AA\ and 0.48 \AA\ for LAE40844 and LAE27878, respectively. Residual bright night sky lines, a problem when extracting faint sources, were interpolated over in each night's averaged spectrum using the SKYINTERP task found in the WMKONSPEC package designed for Keck NIRSPEC reduction\footnote{http://www2.keck.hawaii.edu/inst/nirspec/wmkonspec.html}.  

Flux calibration proceeded using the bright continuum sources that shared the slit with our LAEs as described in the paragraph above. Henceforth these continuum objects will be called calibration stars.   This process corrected for telluric absorption and transformed our flux to F$_{\lambda}$ units.  LAE40844 was calibrated using SDSS J100126.08+021902.2 and LAE27878 was calibrated using SDSS J100025.10+022552.0.  We flux calibrated each night's calibration star spectrum using an appropriate Pickles model spectrum \citep{pic98}, scaled in flux to match the apparent V magnitude of the observed calibration star.  The appropriate Pickles model was chosen based on the spectral type of the calibration star and spectral type was determined from SDSS u-g and g-r colors of the calibration stars as outlined in \citet{fuk10}.  The SDSS u, g and r magnitudes come from SDSS DR7.  The V magnitude of the observed of the calibration star was determined from its SDSS colors and the Lupton (2005) color transformation from SDSS g-r color to V magnitude\footnote{http://www.sdss.org/dr6/algorithms/sdssUBVRITransform.html}.   We then created a sensitivity curve by dividing the scaled down Pickles model by the calibration star's stellar spectrum in counts.  We then multiplied each night's LAE spectrum (in counts) by that night's sensitivity curve to produce a final flux-calibrated LAE spectrum. 

\subsection{Cross check of photometric redshift}
We cross checked the coordinates of each LAE with the sources in the COSMOS Photometric Redshift Catalog Version 1.5 \citep{ics09}.  We found a unique match for each object, within 1$\arcsec$ in all cases.  The photometric redshift of LAE40844 is  $z_{phot} =  3.094$, with a 68\% confidence range of  $3.08 < z_{phot} < 3.11$.  The photometric redshift of LAE27898 is 3.086, with a 68\% confidence range of $3.02 < z_{phot} < 3.12$.  Finally,  the photometric redshift of  LAE14310 is 3.035,  with a 68\% confidence range of $2.98 < z_{phot} < 3.11$.

\subsection{Cross check with Chandra COSMOS  X-ray Sources}
We also compared the locations of our LAEs with the Chandra COSMOS Survey Point Source Catalog \citep{elv09} to exclude contamination from AGNs. The Chandra COSMOS Survey Point Source Catalog contains 1761 X-ray sources in the full 0.5 -- 10keV band with a limiting depth of 5.7$\times10^{-16}$ erg s$^{-1}$ cm$^{-2}$.  The survey covers the central $\sim$ 0.9 deg$^2$ of the COSMOS field.   We find no X-ray sources matching the coordinates of any of our LAEs within 12.8$\arcsec$, which is much larger than the combined positional uncertainties of the narrowband and X-ray catalogs.  This gives upper limits $f_x / f_\lya\la 1.6$ -- $6.0$ for the three sources with LUCIFER spectra--- below the typical ratio $f_x / f_\lya\sim 8$ for type~I quasars, and overlapping the range ($f_x / f_\lya \sim 3$ -- $4$) observed for type~II quasars \citep[e.g.,][]{zhe02}.  Thus the X-ray observations suggest that the \lya\ in these objects is indeed powered by star formation rather than AGN activity, though the present X-ray data are not deep enough to prove this case by themselves.  Also we note that the modest \oiii\ velocity widths of $\sim$ 200 -- 300 \kms\ seen in our two LAEs are much lower than the typical velocity widths of around  1000 \kms\ expected for Type 1 AGN.  

\section{RESULTS}
\label{sec:results}
\subsection{Gaussian Fits to the  [O\,{\sc iii}]  and \lya\ Lines}
To determine the central wavelength and line flux of each emission line we fit a Gaussian plus constant to each emission line.  In the case of the [O\,{\sc iii}] line, we fit a symmetric Gaussian to the line using the MPFITEXPR IDL routine, which is part of the MPFIT package.\footnote{developed by Craig Markwardt http://www.physics.wisc.edu/~craigm/idl/idl.html}  For the \lya\ line we fit an asymmetric Gaussian by modifying the ARM\_ASYMGAUSSFIT IDL routine\footnote{developed by Andrew Marble http://hubble.as.arizona.edu/idl/arm/}, which also utilizes the MPFITEXPR routine.  In its unmodified form, ARM\_ASYMGAUSSFIT basically fits the left and right sides of the central wavelength with different Gaussians and then requires that in the final fit the left and right curves must have the same center and same amplitude where they meet, meaning there are eight parameters, four for each side of the curve (amplitude, center, sigma, constant) but only six of these are free parameters.  This allows for a single curve to be fit, but the curve can have different sigma values for the right and left sides of curve, making it ideal for fitting a \lya\ line with a truncated blue side and extended red wing. We modified ARM\_ASYMGAUSSFIT by fixing the constant on the left side of the \lya\ emission line to a pre-determined constant measured as the average continuum level from 4000 -- 5000 \AA.  This reduces the number of free parameters from 6 to 5 when fitting the \lya\ line.  The constant on the right side of the \lya\ line is allowed to vary, as one can expect a slightly higher continuum level redward of rest-frame \lya. 

We quantified the asymmetry of the \lya\ peaks by defining the ratio of the redside best-fit sigma to the blue side best-fit sigma, or asymmetry$ = \sigma_{red} / \sigma_{blue}$.  Using this definition, any line with an asymmetry measure $>$ 1.0 is considered asymmetric.  From this definition of we find asymmetry measurements of 1.1 $\pm$ 0.1 , 2.1 $\pm$ 0.2  and 1.0  $\pm$ 0.1 for LAE14310, LAE40844 and LAE27878, respectively, meaning the \lya\ line in LAE40844 is highly asymmetric, whereas the \lya\ lines in LAE14310 and LAE27878 appear to be symmetric within the errors.  For comparison with other asymmetry measurements in the literature we also calculated asymmetry using a$_\lambda$ and a$_f$ \citep{rho03} from the best fit asymmetric Gaussians. a$_\lambda$ is 1.2, 2.2 and 1.2 and a$_f$ is 1.2, 1.8 and 1.1 for LAE14310, LAE40844 and LAE27878, respectively.  

 We defined the redshift of the emission line using the central wavelengths determined from these fits (from z $=$($\lambda_{obs}$/$\lambda_{em}$) $- 1$ where $\lambda_{em}$ is the rest-frame vacuum wavelength and $\lambda_{obs}$ is the central wavelength of the best fit).  Line flux for the [O\,{\sc iii}] line was determined from the area under the best fit  symmetric Gaussian. Line flux for the \lya\ line was determined from the narrowband line flux.  The area under the best fit asymmetric Gaussian was scaled to match this flux. The passband of the filter transmission curve was taken into account to assign the appropriate amount of flux to the main \lya\ peak in LAE40844 and this object's secondary 'blue bump' discussed in more detail in section 3.4.  Errors on the area were determined directly from the PERROR output from the MPFITEXPR routine, which returns the one-sigma error on fitted parameters.  PERROR output values were also used to quantify the error on the best fit central wavelength, but an additional error term was included here to account for wavelength calibration errors from the Hectospec and LUCIFER spectra.   Errors on calculated values for redshift and velocity offsets between the \lya\ line and the [O\,{\sc iii}] line were derived from best fit central wavelength and its associated error as described directly above.  Best fits are calculated from unsmoothed spectra, while the spectra in figures 2 and 3 are plotted after 3-pixel boxcar smoothing.

\subsection{[O\,{\sc iii}] Detection with LUCIFER}
We detect the [O\,{\sc iii}] 5008.240/4960.295 \AA\ doublet in two of the three LAEs, LAE40844 and LAE27878.
For the stronger  [O\,{\sc iii}] line (rest frame vacuum wavelength of 5008.240 \AA), we measure a line flux of 35.48 $\pm$ 1.15 $\times$ 10$^{-17}$ erg s$^{-1}$ cm$^{-2}$ in LAE40844 and 6.96 $\pm$ 0.33 $\times$ 10$^{-17}$ erg s$^{-1}$ cm$^{-2}$ in LAE27878. 

The second strongest [O\,{\sc iii}] line (rest frame vacuum wavelength of 4960.295 \AA)  was also found in the same two LAEs.  The line fluxes measured for this line from best fit Gaussians were 14.82 $\pm$ 2.24 $\times$ 10$^{-17}$erg s$^{-1}$ cm$^{-2}$ and 1.47 $\pm$ 0.37 $\times$  10$^{-17}$erg s$^{-1}$ cm$^{-2}$ for LAE40844, LAE27878, respectively.  The ratio of this secondary [O\,{\sc iii}] line to the stronger [O\,{\sc iii}] line is within 2$\sigma$ of the theoretical value ($1/3$) in both galaxies.  This provides a check of the data analysis and increases confidence that this is the 4960.295 \AA\ line. Table 1 summarizes the [O\,{\sc iii}] and \lya\ line fluxes for each LAE, along with relevant broadband and narrowband characteristics. 

LAE14310 showed no detectable  [O\,{\sc iii}] emission. This could be explained if LAE14310 was in fact a lower redshift interloper, but a visual inspection of the optical spectrum has ruled out the object as an [O\,{\sc ii}] emitter at z $\sim$ 0.34 or an [O\,{\sc iii}] emitter at z $\sim$ 0.  In the case of an [O\,{\sc ii}] emitter at z $\sim$ 0.34 we would expect to see [O\,{\sc iii}] at $\lambda \sim$ 6710 \AA, which we do not see.  If the object were a z $\sim$ 0 [O\,{\sc iii}] emitter, we would expect to see the $\lambda =$ 4960.295 \AA [O\,{\sc iii}] line with a line flux of $\sim$ 1.8 $\times$ 10$^{-16}$, which we also don't see.  The more likely scenarios are then that the  [O\,{\sc iii}] emission in this object is weak, or the [O\,{\sc iii}] line is being covered by strong OH lines/H$_2$O absorption in this region. 

We are unable to detect H$\beta$ and/or determine upper limits for H$\beta$ emission, likely because the redshift of each object has placed the H$\beta$ line under strong OH lines and/or under H$_2$O absorption features.  We do not detect the [O\,{\sc ii}] (3729.875 \AA) line by visual inspection in either LAE27878 or LAE40844.  Determining an upper limit for this line by fixing the expected [O\,{\sc ii}] wavelength based on the redshift measured from the [O\,{\sc iii}]  line did not yield a significant upper limit.

\begin{deluxetable*}{lccc}
\tablehead{
\colhead{Galaxy Characteristics} &       \colhead{LAE14310} & \colhead{LAE27878} & \colhead{LAE40844}\\
}
\startdata
u$^{*}$ Magnitude &     25.46 $\pm$ 0.29 &      26.54    $\pm$ 0.50 &   25.56 $\pm$ 0.32 \\
\small{Narrowband KPNO MOSAIC[O\,{\sc iii}] Magnitude} &     22.56   $\pm$ 0.11 &    23.34    $\pm$ 0.15 &   21.82 $\pm$ 0.06  \\
g$^+$ Magnitude    &  24.49    $\pm$ 0.13 &   25.47  $\pm$ 0.19 &    23.66 $\pm$ 0.06\\
\small{z$_{Ly \alpha}$}\tablenotemark{1} & 3.11043 $\pm$ 0.00021 & 3.12051 $\pm$ 0.00021  & 3.11639 $\pm$ 0.00021\\
\small{z$_{OIII}$} & & 3.11879 $\pm$ 0.00011 & 3.11170 $\pm$ 0.00014\\
\small{\lya\ Equivalent Width}\tablenotemark{2}$^{,6}$  & 89$^{+17}_{-20}$ & 118$^{+34}_{-40}$  & 78$^{+8}_{-8}$\\
\small{\lya\ Line Flux from Narrowband}\tablenotemark{3}$^{,4}$  & 18.7$^{+2.25}_{-2.51}$  & 9.41$^{+1.42}_{-1.63}$ & 36.1$^{+2.35}_{-2.47}$ \\
\small{Upper limit on xray / \lya\ Flux Ratio} & 3.0  & 6.1 & 1.6 \\
\small{O\,{\sc iii} line flux ($\lambda$ = 5008.240 \AA)}\tablenotemark{3}  & & 6.96 $\pm$ 0.33 & 35.48 $\pm$ 1.15  \\
\small{O\,{\sc iii} line flux ($\lambda$ = 4960.295 \AA)}\tablenotemark{3}  & & 1.47 $\pm$ 0.37 & 14.82 $\pm$ 2.24 \\
\small{O\,{\sc iii} velocity width($\lambda$ = 5008.240 \AA)}\tablenotemark{5} & &189.3 $\pm$ 10.3 & 281.1 $\pm$ 9.8 \\
\small{O\,{\sc iii} FWHM ($\lambda$ = 5008.240 \AA)}\tablenotemark{6} & &13.0 $\pm$ 0.7 & 19.3 $\pm$ 0.7 \\
\small{v$_{\textrm{offset}}$ of O\,{\sc iii} from Ly$\alpha$}\tablenotemark{5}& & +125 $\pm$ 17.3 &  +342 $\pm$ 18.3 \\
\enddata
\tablenotetext{1}{corrected for Earth's motion \hspace{0.3cm}}
\tablenotetext{2}{Rest Frame, from Narrowband flux, calculated as $(F_{NB} - F_g)/(F_g/55\AA - F_{NB}/1265\AA)$ where 55 \AA\ is bandpass of KPNO MOSAIC[O\,{\sc iii}] filter and 1265 \AA\ is bandpass of Subaru g$^+$ filter, F$_{NB}$ is flux in narrowband, F$_{g}$ is flux in g$^+$ band. \hspace{0.3cm}}
\tablenotetext{3}{10$^{-17}$ erg s$^{-1}$cm$^{-2}$ \hspace{0.3cm}}
\tablenotetext{4}{Calculated as $(F_{NB} - F_g)(c / \lambda_c^2)d\lambda$ where $\lambda_c$ central wavelength  and d$\lambda$ is bandpass of KPNO MOSAIC[O\,{\sc iii}] filter \hspace{0.3cm}}
\tablenotetext{5}{\kms \hspace{0.3cm}}
\tablenotetext{6}{\AA \hspace{0.3cm}}
\end{deluxetable*}

\subsection{Systemic Redshifts and Velocity Offsets Between [O\,{\sc iii}] and \lya}
Using the Gaussian fits described above, we measured systemic redshifts from the  strongest [O\,{\sc iii}] line in the two objects with detections, finding redshifts of 3.11170 $\pm$ 0.00014 and 3.11879 $\pm$ 0.00011 for LAE40844 and LAE27878, respectively.

Measuring the redshift of each object using the \lya\ line instead of the [O\,{\sc iii}], yields redshifts of 3.11639 $\pm$ 0.00021 and  3.12051  $\pm$ 0.00021  for LAE40844 and LAE27878, respectively, after corrections for the Earth's motion.  To correct for the Earth's motion we calculated topocentric radial velocities\footnote{http://fuse.pha.jhu.edu/support/tools/vlsr.html} for the two observing locations (MMT and LBT) for the nights the objects were observed at each location.  The generally accepted interpretation of this discrepancy in redshift measurements from interstellar emission lines and \lya\ is that there is a kinematic offset between the lines caused by a large scale outflow, an outflow likely driven by active star formation.  

Assuming that the emission lines originate from a single redshift, we pin the lines to the redshift of the [O\,{\sc iii}] line and use this frame to calculate a velocity offset between the two lines.  We justify using the nebular emission to define the systemic redshift of the galaxy since the [O\,{\sc iii}] emission originates from H\,{\sc ii} regions surrounding ionizing stars.  These regions ought to be at the systemic redshift of the galaxy.  While the \lya\ initially departs from the same regions, resonant scattering, which effects \lya\ and not [O\,{\sc iii}], changes the observed location of \lya\ emission.   

We derived velocity offset between the 5008.240 \AA\ [O\,{\sc iii}] line and \lya\ line based on the central wavelength of each line, determined by the best fit asymmetric and symmetric Gaussians for the \lya\ and [O\,{\sc iii}] lines, respectively.  We find velocity offsets of +342 $\pm$ 18.3 km s$^{-1}$ and +125 $\pm$ 17.3 km s$^{-1}$ for LAE40844 and LAE27878, respectively.  The velocity offsets between the [O\,{\sc iii}] and \lya\ lines are shown in Figure 2.

Steidel et al. (2010) note that their redshift determinations for z $\simeq$ 2 -- 3 galaxies based on NIR H$\alpha$ measurements have an inherent uncertainty of $\sim$ 60 \kms.  This estimate is based on repeated observations of the same galaxy with their 0.76\arcsec\ slit in different positions.  The uncertainty is explained as arising from the fact that in each measurement they are only measuring the velocity of the fraction of the flux that entered the slit.  We find that our [O\,{\sc iii}] measurements for our z $\sim$ 3.1 galaxies should not be subject to such a large uncertainty from this effect due to our larger slit width (1\arcsec) and small galaxy sizes.  The sizes of our galaxies, from ACS i-band half light radii, are 1.1 and 1.3 kpc, for LAE40844, LAE27878, respectively \citep{ma10}.  The corresponding half-light angular diameters are still less than half the slit width.  The 1.5\arcsec\ diameter fibers should also minimize flux losses for our \lya\ observations, however, we concede that our error bars may be lower limits due to such systematics we may not be fully taking into account.

\begin{figure}
\epsscale{1.0}
\plottwo{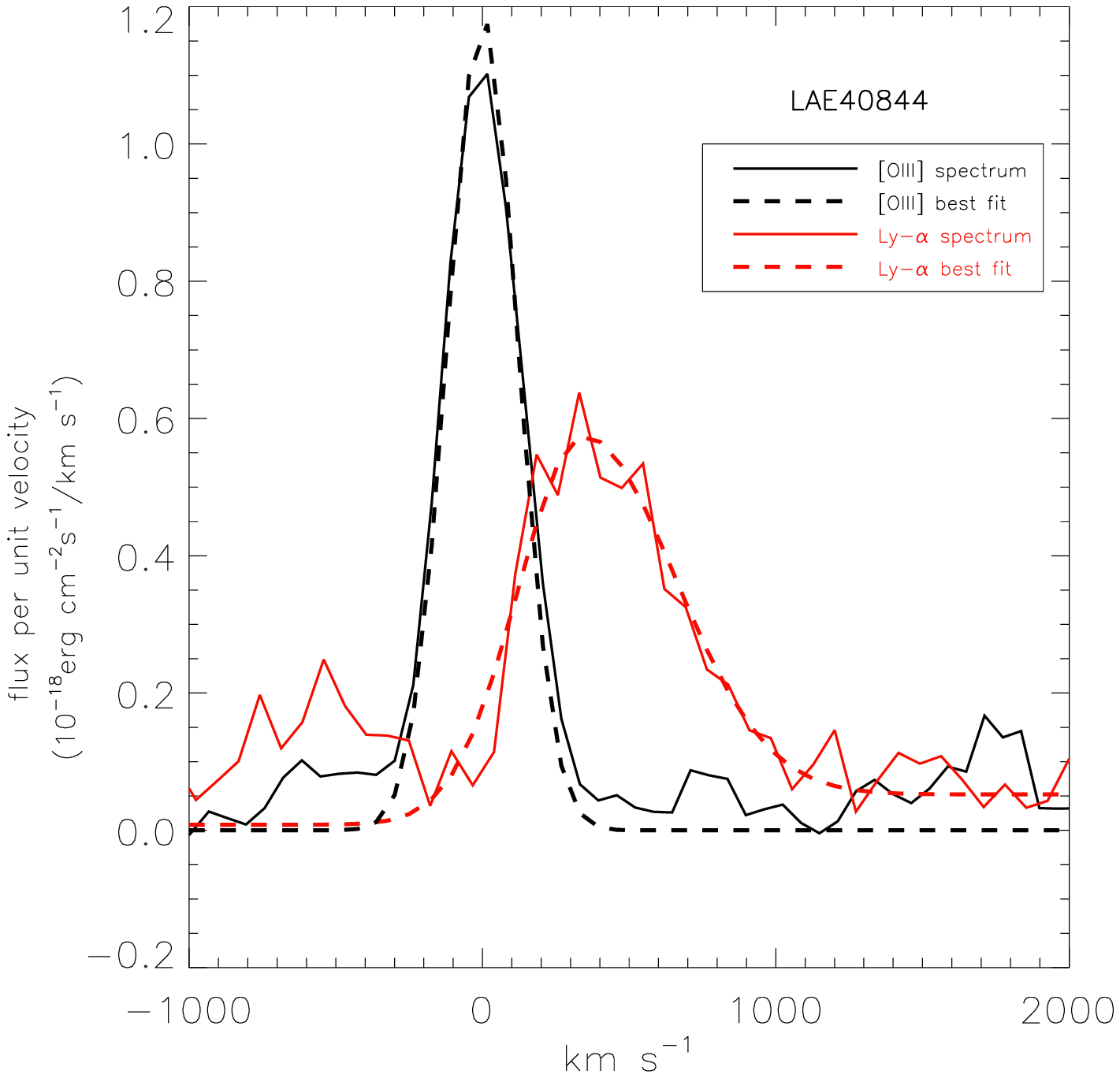}{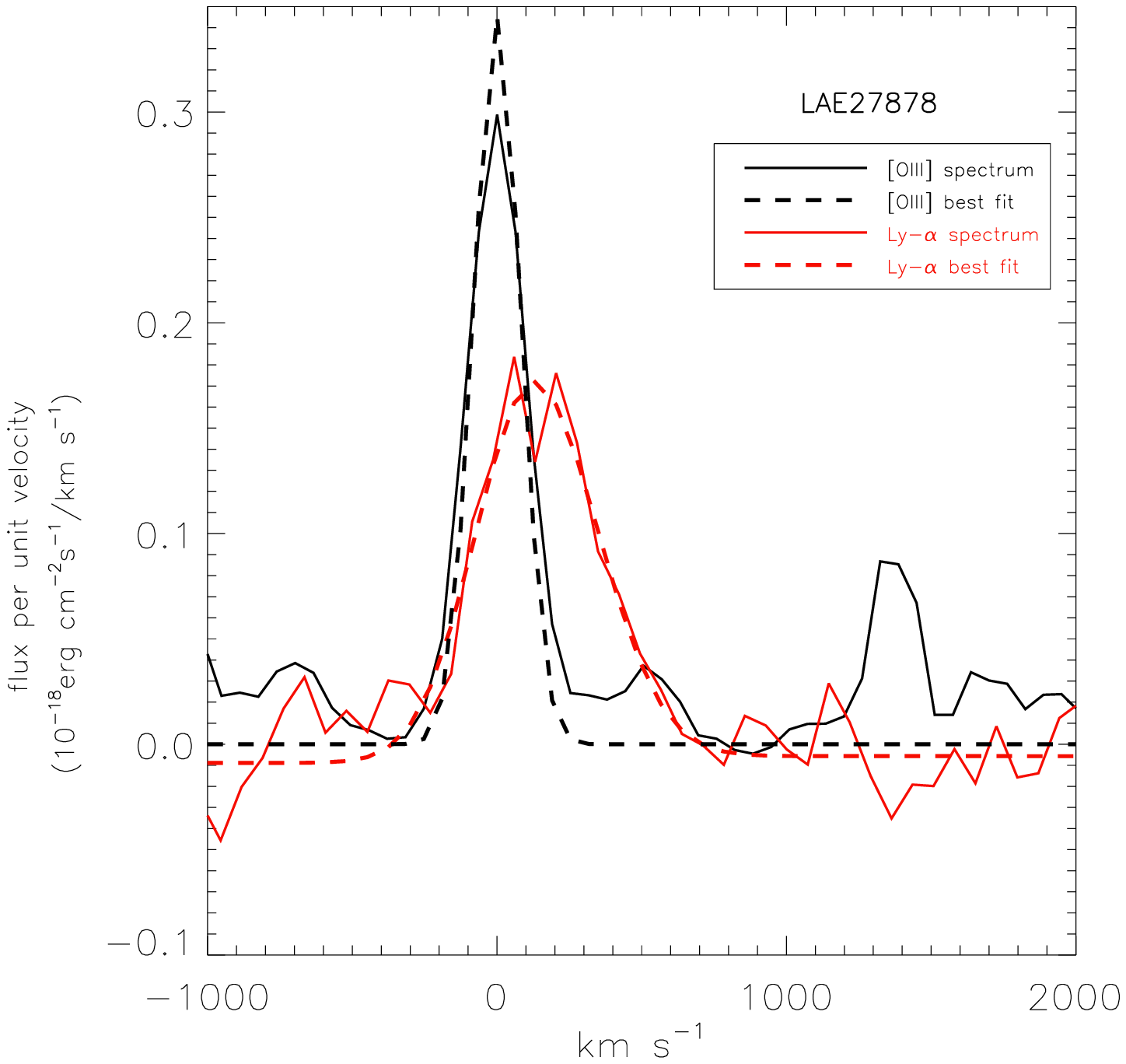}
\caption{[O\,{\sc iii}]  line and \lya\ lines with their corresponding best-fit Gaussians.  Velocity offset between [O\,{\sc iii}] and \lya\ line for LAE40844 is 342 km s$^{-1}$. Velocity offset between [O\,{\sc iii}] and \lya\ line for LAE27878 is 125 km s$^{-1}$.  Optical spectra have been calibrated using the \lya\ line flux determined from narrowband imaging.  The feature near +1400 \kms\ in LAE27878 is a residual night sky line at $\sim$ 20728.17 / 20729.859 \AA\ \citep{rou00}.}
\end{figure}

\subsection{`Blue bump' in LAE40844 - Velocity Offset of Secondary \lya\ Feature}
LAE40844 has another feature of interest in its optical spectrum, namely a smaller, secondary \lya\ peak blueward of the systemic velocity of the object.  See Figure 3 for a detailed view of this feature.  This feature is fit with an asymmetric Gaussian as described for the main \lya\ line in section 3.1, but the constants on the left and right sides of the Gaussian are required to be equal (and to be equal to the pre-determined constant level also described in section 3.1) to ensure that the main \lya\ peak did not interfer with our best fit measurements of this secondary peak.  This essentially reduces the number of fitted parameters for the blue bump from 5 to 4, meaning that when both the main \lya\ line and the blue bump are fit, a total of 9 parameters are returned (5 for main \lya\ peak, 4 for blue bump).  Our method yields a velocity offset from the [O\,{\sc iii}] line of -453.7 $\pm$ 50.7 \kms, after correction for the Earth's motion.  From this measurement we determine that the two \lya\ peaks are offset from one another by +796.2 $\pm$ 53.9 km s$^{-1}$.

Additionally, using the flux calibration we derived from the narrowband line flux, we find that this blue \lya\ peak has a line flux of $\sim$ 1.08 $\times 10^{-16}$ erg s$^{-1}$ cm$^{-2}$.  When determining this calibration we found that based on the transmission curve of the narrowband filter, the blue bump contributed $\sim$ 9.4 \% of the total narrowband line flux (3.61 $\times 10^{-16}$ erg s$^{-1}$ cm$^{-2}$).  Comparing this to a line flux of $\sim$ 3.27 $\times 10^{-16}$ erg s$^{-1}$ cm$^{-2}$ for the red \lya\ peak we find an approximate flux ratio, red:blue, for the two lines of 3.0.  In other words, the strength of secondary (blue) peak is roughly 33\% that of the main (red) peak.  In section~\ref{sec:discuss} we discuss a scenario that can give rise to this blue bump and compare the velocity offset we find between the two \lya\ lines to velocity offsets that have been presented in the literature on \lya\ radiative transfer. 

\begin{figure}
\epsscale{0.6}
\plotone{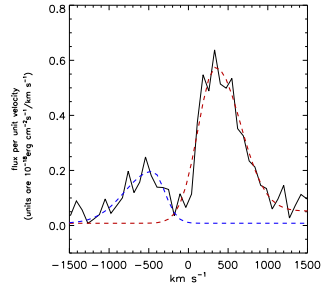}
\caption{LAE40844, best fit asymmetric Gaussian to redshifted \lya\ emission in red, best fit asymmetric Gaussian to blue shifted \lya\ emission in blue, optical spectrum in black.  Velocity offset between the \lya\ two peaks is +796.2 $\pm$ 53.9 km s$^{-1}$.}
\end{figure}

\subsection{Effect of the [O\,{\sc iii}] Emission Line on Mass and Age Estimates}
The COSMOS field has a deep K$_{s}$-band (centered at 21460\AA, $\Delta\lambda$ = 3250 \AA) coverage fromCFHT WIRCAM, with a $5\sigma$ depth of $\sim$ 23.8 mag in a 3$\arcsec$ aperture on a PSF matched image.  For our z $\sim$ 3.1 LAEs, this band will encompass both the continuum and [O\,{\sc iii}] emission.  The 2008 COSMOS Intermediate and Broad Band Photometry Catalog \citep{cap07} has K$_{s}$ band magnitudes of 22.61 $\pm$ 0.07 for LAE40844, and 24.91 $\pm$ 0.62 for LAE27878 (MAG\_AUTO measurements from a 3$\arcsec$ aperture).  Our measured [O\,{\sc iii}] line fluxes can account for the entire K$_s$ band fluxes, where we find K$_s$-band magnitudes of 23.09 $\pm$ 0.036 and 24.79 $\pm$ 0.053 for LAE40844 and LAE27878, respectively, using just the [O\,{\sc iii}] line fluxes.  {\it This shows that our detected [O\,{\sc iii}] emission lines alone can be responsible for all or nearly all the flux measured in the K$_{s}$ band for both of these LAEs.}

This has important implications for mass and age estimates of high redshift galaxies.  These estimates typically rely on the size of the Balmer break to determine the  age of the galaxy.  If, as is the case for our two LAEs, there is a dominant emission line polluting the location redward of the Balmer / 4000 \AA\ break, then the size of the break may be overestimated and the subsequently derived ages and stellar masses may be overestimated.  \citet{sch09} found that when nebular emission lines were included, SED fitting of the \citet{ey07} sample of 10 z $\sim$ 6 galaxies yielded an average age $\sim$ 4 times younger than what was found without the emission lines included.  The average stellar mass estimate also decreased from 1.2$\times$10$^{10}$ \msol\ without emission lines to 7.8$\times$10$^{9}$ \msol\ when nebular emission was included.   Some studies of \lya\ emission in Lyman-break selected populations have found that the strongest \lya\ emitters have blue UV spectral slopes but red optical slopes \citep{shap03,korn10}, and have concluded that \lya\ emission is strongest in LBGs that are older but relatively dust-free.  If strong nebular line emission contributes to the observations of red rest-optical slope, it might be possible to reinterpret such observations in terms of young, strongly line emitting galaxies, although \citet{korn10} argue that their observed correlations between \lya\ strength and stellar population age are unchanged when they select objects only at redshifts where the optical continuum filters are line-free.  Previous work on stellar populations of \lya-selected galaxies has found that nebular line emission is required to explain observed rest-optical colors \citep{f08}.  Recently, at the highest redshifts, \citet{ono10} have shown that either old stellar populations or young ones with strong nebular emission can reproduce the composite SEDs of \lya\ selected galaxies.  The older models have correspondingly higher stellar masses, since mass to light ratio increases strongly with age. Our observations provide {\it direct} observational evidence that nebular line flux dominates the rest optical in analogous objects at $z\approx 3$, and hence supports the interpretation of the high redshift \lya\ selected populations as young and low-mass objects.

\section{DISCUSSION} 
\label{sec:discuss}

\subsection{[O\,{\sc iii}] Luminosities} 
The [O\,{\sc iii}] line has been measured in other objects at similar redshifts.  Pettini et al. (2001, henceforth P01) find  [O\,{\sc iii}] (5007 \AA) luminosities of 3.5 -- 15.6 $\times$ 10$^{42}$ erg s$^{-1}$ in four individual z $\sim$ 3.1 LBGs observed with VLT1/ISAAC and/or Keck II/NIRSPEC. Comparing this to our range of luminosities for the 5007 \AA\ line for two z $\sim$ 3.1 LAEs of 6.1 -- 31.0 $\times$ 10$^{42}$ erg s$^{-1}$, it appears that our fainter 5007 \AA\ measurement falls in the P01 range, while the stronger of our two 5007 \AA\ lines  is almost double that of the brightest luminosity in the P01 sample. Luminosities for the 4959 \AA\ line in the P01 sample are 1.4 -- 6.5 $\times$ 10$^{42}$ erg s$^{-1}$.   This yields the same trend we see in the 5007 \AA\ line; where our fainter 4959 \AA\  measurement falls in the P01 range, and our brighter 4959 \AA\ line is approximately twice that of the brightest 4959 \AA\ measurement in the P01 sample.

[O\,{\sc iii}] luminosities from lensed galaxies around z $\sim$ 3 have also been documented. \citet{fos03} measured the [O\,{\sc iii}] line in a lensed H\,{\sc ii} galaxy, also known as the Lynx arc \citep{ho01}, at a redshift of z $\sim$ 3.36.  Using NIRSPEC K-band spectra they find an  [O\,{\sc iii}] luminosity of 28.3 $\pm$ 0.3 $\times$ 10$^{42}$ erg s$^{-1}$ for the 5007 \AA\ line and 9.8 $\pm$ 0.3 $\times$ 10$^{42}$ erg s$^{-1}$ for the 4959 \AA\ line.  These luminosities (which have been corrected for magnification) are quite comparable to our measurements for the 5007 \AA\ and 4959 \AA\ lines.  \citet{f09b} measured the [O\,{\sc iii}] line in a lensed ultraviolet-luminous z = 2.73 galaxy known as the 8 o'clock arc.  For the 4959 \AA\ line they find a luminosity of 8.9 $\pm$ 0.4 $\times$ 10$^{42}$ erg s$^{-1}$ (after correction for magnification).  This again falls right in the range reported for our two LAEs.  The 5007 \AA\ line was not reported by \citet{f09b} because it fell in an area of low atmospheric transmission which required a correspondingly large tellluric correction and led to large uncertainties in any measurements from the line.

Looking at our sample of two galaxies, the \oiii\ luminosities in LAE27878 are most similar to the P01 LBGs while the more luminous \oiii\ lines in LAE40844 are more comparable to more luminous \oiii\ lines found in galaxies studied via lensing. A larger sample of NIR spectra with \oiii\ line measurements for LAEs will help us understand the range of \oiii\ luminosity in LAEs and its implications.

\subsection{\lya\ Line Profiles and Outflow Models} 
In addition to the information that can be gleaned from the line fluxes of nebular emission lines such as [O\,{\sc iii}], the asymmetric profiles of \lya\ emission lines themselves carry information on the physical conditions and processes in these objects.  Resonant scattering can lead to asymmetric profiles through radiative transfer 
processes operating either within the \lya\ emitting galaxy, or in the surrounding intergalactic medium.  Establishing the systemic velocity with the \oiii\ line sheds new light on those processes and conditions.

We find that the \lya\ line profiles seen in  LAE40844 and LAE27878, and the velocity offset of the \lya\ line from systemic in both objects is in good agreement with what is predicted by outflow models, where the \lya\ is redshifted through interaction with receding gas on the far side of the galaxy, and transmitted through approaching gas on the near side due to the line's kinematic redshift. Two particular types of outflow models are discussed here due to their apparent agreement with our results - the case where the outflow is in a coherent shell \citep{tt99,daw02,zhe02,v06,v08} and the case of a clumpy outflow \citep{neu91,ho06,stei10}.

Understanding our results in the context of an outflow is justified in that high-redshift LAEs are typically young (age $\simeq 10^7$ years) with vigorous star formation \citep[e.g.,][]{f07,f09,pir07,gaw07,lai08}.  Their typical star formation intensities are well above the threshold required to drive galactic winds \citep{ma10}.  Similar winds are seen in nearby starbursting galaxies, with velocities of order 10$^2$-10$^3$ km s$^{-1}$ \citep{he90,he02}, numbers that encompass our measured velocity offsets.

In the model with a single expanding shell, with a central monochromatic source  (Verhamme et al. 2006, henceforth V06), the redshifted \lya\ line is built up of photons that underwent one or more backscatterings off the expanding shell.  The more backscatterings the photon undergoes, the further it is redshifted, giving rise to the prominent red wing that is seen in the redshifted \lya\ line.  Photons that are emitted from the blue wing of the \lya\ line in the part of the shell that is approaching the observer can give rise to the blue bump we see in LAE40844.  See Figure 12 of V06 for a detailed description of photons escaping from an expanding shell.

For the parameter space examined in the V06 simulations, velocity offsets of a few 100 \kms\ are predicted for the redshifted \lya\ line.  The magnitude of the \lya\ velocity offset depends on the velocity of the expanding shell, the column density of neutral hydrogen and the Doppler parameter (see V06 for details on these parameters).  Our velocity offsets of 125 \kms\ and 342 \kms\ appear to be at the low end of this range.  When the blue bump is seen in these simulations it is offset from the redshifted \lya\ line by $\sim$ 1000 \kms, in agreement with our measured offset in LAE40844 between the red and blue peaks of 796 \kms.  Thus there is good general agreement between our observations and the V06 models, which have not been tuned specifically to fit our data.  It seems likely that an expanding shell model could fit our data well with some adjustment of the input parameters.  Additionally, while our total offset between the blue and red \lya\ peaks in LAE40844 agree with the total offsets seen in their simulations, we find that v$_{blue\,peak} \sim$ -2v$_{red\,peak}$ in LAE40844.  In V06 the velocity offset between these two peaks is nearly symmetric in the cases where both peaks are present and the expansion velocity of the shell is small ($< 200$ \kms).  Where the expansion velocity of the shell is large (300 -- 400 \kms) the blue peak is nearly -1/2 the velocity shift of the red peak.  In either case, these predictions do not directly match our observations presented here. Further work is needed to understand the discrepancy between the velocities of the redshifted and blueshifted peaks in models and observations. Deviations from spherical symmetry of the expanding shell model could help account for this difference, as was noted in \citet{sv08}.

These same authors further investigate their model in a later paper (Verhamme
et al. 2008, henceforth V08) by applying it to fit the observed \lya\ emission
line profiles of the LBGs from Tapken et al. (2007), which lie
at $2.7 \le z \le 5$.
Varying model parameters include the Doppler parameter, V$_{exp}$,
the neutral hydrogen column density N$_{HI}$,
the dust optical depth $\tau_{a}$, and the intrinisc EW
and FWHM of the input \lya\ line.  Nine of the 11 LBGs investigated were found to
have V$_{exp}$ $\sim$ 150 -- 200 \kms.  Two LBGs, which resemble our LAE40844 in
that they have a blue bump, can be fit
either with larger velocities, of order 300 -- 400 \kms\, or a with a
quasi-static medium with V$_{exp}$ of order 10 -- 25 \kms.
V08 prefer the quasi-static explanation.  We note that our results of
v$_{offset}$ = 125 and 342 \kms\ is fully consistent with the V08 results.
However, we emphasize that we measure the velocity offset
between the \lya\ line and the systemic redshift as defined
by \oiii, whereas V06 and V08 are quantifying V$_{exp}$, which is the velocity
of a spherically expanding shell around the central \lya\ source in their model.
We also note that LAE40844, with its blue bump and
v$_{offset}$ of 342 \kms, appears to show that a quasi-static model is
not always preferred for objects with a blue bump.  

Steidel et al. (2010, henceforth S10) consider outflows to be an important component of the mechanism that shapes the observed \lya\ profile, but they prefer a scenario in which the structure and kinematics of the circum-galactic medium can produce our observed profiles, instead of relying on \lya\ radiative transfer in an expanding shell to create the profiles we see. In the S10 scenario, a clumpy outflowing gas will allow some \lya\ photons to escape from a galaxy producing the redshifted \lya\ line we have observed.  This same scenario is also capable of producing the faint `blue bump' we have discussed in section 3.4.  S10 find an average velocity offset of 445 \kms\ in a sample of 42 z $\simeq$ 2 - 3 LBGs with \lya\ emission.  The velocity offset measurement in this case was made with respect to the redshift defined by H$\alpha$ emission.  As is the case when we compare our results to the Verhamme model, our velocity offsets of 125 \kms\ and 342 \kms\ for our observed LAEs are below these values.  This conclusion includes consideration of the fact that S10 measures the centroid of a single function whereas we measure the peak of Ly-alpha through an asymmetric function composed of two Gauss functions; which inherently causes the S10 measurements to be higher. This causes a 10 -- 15\% difference in the velocity offsets calculated, and hence we stil find our velocity offset values to be lower than those in the S10 sample evn after this consideration.

At present, our sample of two z $\sim$ 3.1 LAEs reported in this paper does not allow us to distinguish between the expanding shell scenario or the S10 interpretation, as both are able to produce profiles and velocity offsets in reasonable agreement with our observations.  Whichever of the scenarios discussed here (V06 or S10) is producing the observed velocity offsets, we emphasize that winds/outflows are important in either case.

The additional observational samples to which we can best compare our current results are z $\sim$ 3 LBGs.  P01 and Shapley et al. (2003, henceforth S03) have both measured the velocity offset of the \lya\ line from systemic in a population of z $\sim$ 3 LBGs. P01 find velocity offsets for the \lya\ line of 200 - 1100 \kms\ in a sample of 13 LBGs that also show \lya\ in emission.  The velocity offset is measured compared to the redshift of nebular H{\sc ii} emission.  S03 find a velocity offset of 360 \kms\ from a composite spectrum of 811 z $\sim$ 3 LBGs.  The \lya\ velocity offset measurement in S03 was made with respect to interstellar absorption lines.  \citep{tap07} measured a velocity offset between LIS lines and \lya\ in seven LBGs with \lya\ emission at redshifts of 2.7 -- 5 (the same sample of galaxies used in V08) finding an average offset of 580 \kms. Including the 445 \kms\ offset measurement from the S10 sample discussed above, we note that in all cases our observed velocity offsets of 125 - 342 \kms\ in two z  $\sim$ 3.1 LAEs are at the low end of the values reported for the various LBG samples, although the velocity offset measurements have thus far been made via different methods.

An alternative explanation of the observed \lya\ profiles, not based on galactic scale outflows, is that they arise through resonant scattering in the intergalactic gas surrounding an LAE.  Zheng et al. (2010) have explored such a mechanism in detail for redshift $z=5.7$.  Their models can produce \lya\ lines that qualitatively resemble our observations both in the line asymmetry and in the redshift of the \lya\ line.    However, some caution is needed in applying these results to our data set, given that the IGM density at redshift $z=3.14$ is $\sim 1/4$ that at $z=5.7$, and the ratio of neutral gas density between these two redshifts is still more extreme.  
Overall, we consider it more likely that winds play an important role in \lya\ escape, given that winds are  generically expected for galaxies with the high specific star formation rates typical of LAEs.

\subsection{Implications of Detected Outflows} 
Detection and characterization of galaxy scale outflows at high redshift is important because these outflows have important consequences for the evolution of individual galaxies as well as the evolution of the IGM.  Large scale galactic outflows are capable of driving materials out of the galaxy and may therefore contribute to metal enrichment of the IGM at high-z by introducing materials produced from starbursts into the IGM.  In addition, galactic winds likely provide a crucial channel by which ionizing photons can escape from a galaxy \citep{stei01,he01}.  This has important implications for the contribution of high-z galaxies to the reionization of the IGM.   In terms of shaping an individual galaxy, superwinds are responsible for driving dust from a starbursting galaxy \citep{ferr91,he00,shap,bf05} and the mass loss from a galaxy due to an outflow may be capable of suppressing star formation \citep{sp99,he02}.  While our work presented in this paper has now demonstrated that that our sample of two \lya\ selected galaxies at z $\sim$ 3.1 are driving winds, further characterization of these winds from a larger sample will help us understand and test some of the broader implications of winds detailed above.

The observed velocity offsets between the \lya\ line and systemic velocity also have important implications for \lya\ based tests of reionization \citep{mr04,san04,mr06,dw10}. In particular, the ionized volume test proposed by \citet{mr06} (hereafter MR06)
is much more sensitive if $\Delta v$ (\lya\ peak vs. systemic) is typically small.   That test works by 
noting that substantial transmission of \lya\ through a generally neutral IGM requires a locally ionized region around each observed LAE.  The product of the bubble volume $V$ and the LAE number density $n$ is then a filling factor of ionized gas, from which the volume fraction of the ionized phase 
is $\approx 1 - exp(-nV)$ (MR06).  The relevant bubble volume $V$ is sensitive to the velocity offset,
since \lya\ photons that are already redshifted before leaving the emitting galaxy are less strongly scattered by the damping wing of neutral hydrogen in the surrounding intergalactic gas.  The effect is explored in detail in Figure 1 of MR06.  Replacing the range  $0 \le \Delta v \le  360$ \kms from that
paper with our measurements, $ \Delta v = 125$ \kms and $\Delta v = 342$ \kms, would narrow the range
of permitted volume ionized fractions from the 20\% -- 50\% range derived in MR06 to $\sim$ 20\% -- 40\%.
While this discussion is subject to refinement as the sample of LAEs with a measured velocity
offset grows, it shows the importance of measuring $\Delta v$ for studying reionization with 
\lya\ lines.

Finally, even at redshifts where the IGM is predominantly ionized and
affects the \lya\ line only through the \lya\ forest, our systemic
redshift measurements have important implications.  Several groups
have shown that the observed \lya\ luminosity function is largely
unchanged from $z\approx 3$ to $z\approx 6$
\citep[e.g.,][]{daw04,s05,ou08,cas11}.  Recently, \citet{cas11} have
combined this observation with the expected optical depth evolution of
the \lya\ forest \citep{mad,fan06} to infer that the \lya\ luminosity
function is in fact evolving towards higher luminosities at higher
redshifts.  The key assumption in this argument is that the line
emitted by the LAE is symmetric and centered on the systemic velocity,
so that the fraction scattered by the IGM approaches 50\% by $z\approx
6$.  This implies that the fraction of \lya\ flux observed would
decline by a factor of $\approx 0.6$ between redshift $z\approx 3$ and
$z\approx 6$.
However, for the two objects where we observe \oiii, we know that only
those photons observed blueward of the systemic velocity would be
subject to additional \lya\ forest absorption at higher redshift.  For
LAE40844, the blue bump would be progressively obscured by the forest
at higher redshift, resulting in a flux loss of a factor $\approx
3.0/4.0 = 0.75$ or so at most.  For LAE27878, there is no significant
flux blueward of the systemic velocity, and the fraction of its \lya\
emission that we can see should remain nearly unchanged from $z=3.1$
until the IGM neutral fraction becomes so large that the red damping
wing of the IGM becomes optically thick--- i.e., until we reach the
central stages of reionization.

To recap, our objects show relatively little \lya\ flux blueward of
the systemic velocity.  This either requires dropping the assumption
that \lya\ is symmetric and centered at the systematic velocity, or
else implies that we happen to have observed two sources where the
\lya\ forest at $z\approx 3.1$ is unusually optically thick.  We now
proceed to evaluate the likelihood of the second scenario.

Presently, our sample consists of two galaxies, with measured blue to
red flux ratios of 0.33 and 0.08 for objects 40844 and 27878,
respectively.  We estimate the chance of such an occurrence under the
assumption of Cassata et al (i.e., that the intrinsic ratio is 1:1,
and deviations indicate absorption by the \lya\ forest).
\citet{mcd00} give probability distributions for transmission through
the Lyman $\alpha$ forest at $z=3.00$ and $z=3.89$.  Interpolating
their results to our redshift ($z=3.12$), we infer probabilities $P(T
\le 0.33) = 0.24$ and $P(T \le 0.08) = 0.15$.  Given a sample of 2,
and treating their \lya\ forest transmissions as independent random
variables, the probability that at least one will have $T \le 0.08$
while the second has $T \le 0.33$ becomes $0.050$.  So, our
present results disfavor this assumption of an intrinsic 1:1 ratio of
blue:red flux, suggesting that the luminosity function evolution
inferred in \citet{cas11} is a consequence of their implicit
assumption $\Delta v = 0$ and not a strong conclusion about the true
evolution of \lya\ galaxy populations.  For now, this is a $2\sigma$
result. Observations of a few more \lya\ emitters with systemic
velocity measurements could resolve this question firmly.

\section{CONCLUSIONS}
We have detected [O\,{\sc iii}] emission in two \lya\ selected galaxies at z $\sim$ 3.1 using the new NIR spectrograph LUCIFER on the LBT.  This is a successful demonstration that the [O\,{\sc iii}] line can be detected in high-z \lya\ selected galaxies and that this line can be used to investigate the characteristics of these galaxies.

In both LAEs we measured a velocity offset between the \lya\ emission and the systemic redshift of the galaxy as defined by the [O\,{\sc iii}] emission.  These velocity offsets range from 125 - 342 \kms.  We find that these velocity offsets and the observed profile of the \lya\ line both indicate that our measurements are the result of \lya\ emission emerging in the presence of a galactic outflow.  In addition we have measured \lya\ flux blueward of systemic in a `blue bump' in one of our objects.  This is another phenomenon one can expect when observing \lya\ emission in the presence of an outflow.  We find that a scenario in which radiative transfer effects of \lya\ emission emerging from an expanding shell (V06) is able to reproduce reasonably well our observed \lya\ profiles and velocity offsets.  We also find that a scenario in which \lya\ photons escape from a circumgalactic gas as described by \citet{stei10} is capable of reproducing our results reasonably well.

A larger sample of measured velocity offsets will better constrain the range of velocity offsets we can expect for z $\sim$ 3.1 LAEs and will allow for better understanding of how these offsets compare to those observed in Lyman-break selected samples.  These comparisons should shed light on the relationship between crucial characteristics like galaxy mass, star formation rates and the magnitude of the observed velocity offsets.  Finally, larger samples of velocity offsets will further improve our ability to infer constraints on cosmological reionization from observations of \lya\ galaxies at high redshifts.

\acknowledgements We thank the LUCIFER Instrument team and the LUCIFER SDT team for their hard work and for exciting science with this new instrument. This work was supported by NSF grant AST-0808165.  In addition we thank the referree for insightful comments that have improved this paper.


\begin{thebibliography}{96}
\expandafter\ifx\csname natexlab\endcsname\relax\def\natexlab#1{#1}\fi
\bibitem[Ageorges et al.(2010)]{ag10} Ageorges, N. et al. 2010, ``LUCIFER1 commissioning at the LBT'', SPIE Proc. Vol. 7735-56,  to be published
\bibitem[Bertin \& Arnouts(1996)]{ba96} Bertin, E. \& Arnouts, S. 1996, A\&AS, 117, 393
\bibitem[Adelberger et al.(2005)]{ad05} Adelberger, K. L., Steidel, C. C.,Pettini, M., Shapley, A. E., Reddy, N. A. \&   Erb, D. K. 2005, \apj, 619, 697
\bibitem[Bianchi \& Ferrara(2005)]{bf05} Bianchi, S. \& Ferrara, A. 2005, \mnras, 358, 379
\bibitem[Bond et al.(2009)]{bon09} Bond, N. A., Gawiser, E., Gronwall, C., Ciardullo, R., Altmann, M. \& Schawinski, K. 2009, \apj, 705, 639
\bibitem[Bond et al.(2010)]{bon10} Bond, N. A.,  Feldmeier, J. J., Matkovich, A., Gronwall, C., Ciardullo, R., \& Gawiser, E. 2010, \apj, accepted
\bibitem[Bremer \& Johnstone(1995)]{bj95}Bremer, M. N. \& Johnstone, R. M. 1995, MNRAS, 277, L51
\bibitem[Capak et al.(2007)]{cap07} Capak, P. et al. 2007, ApJS, 172, 99
\bibitem[Cassata et al.(2011)]{cas11} Cassata, P. et al. 2011, A\&A, 525, 143
\bibitem[Cowie et al.(2010)]{co10}Cowie, L. L. et al. 2010, ApJ, 711, 928
\bibitem[Dawson et al.(2002)]{daw02} Dawson, S., Spinrad, H., Stern, D., Dey, A., van Breugel, W., de Bries, W. \& Reuland, M.,  2002, \apj, 570, 92
\bibitem[Dawson et al.(2004)]{daw04} Dawson, S., Rhoads, J. E., Malhotra, S., Stern, D., Dey, A., Spinrad, H., Jannuzi, B. T., Wang, J. X. \& Landes, E. 2004, \apj, 617, 707
\bibitem[Dayal et al.(2010)]{day10} Dayal, P., Maselli, A. \& Ferrara A. 2010, MNRAS, 410, 830
\bibitem[Dijkstra \& Wyithe(2010)]{dw10} Dijkstra, M. \& Wyithe, S. 2010, MNRAS, 408, 352
\bibitem[Dow-Hygelund et al.(2007)]{dh07} Dow-Hygelund, C. C. et al. 2007, \apj, 660, 47
\bibitem[Elvis et al.(2009)]{elv09} Elvis, M. et al. 2010, APJS, 184, 158
\bibitem[Eyles et al.(2007)]{ey07} Eyles, L. P. et al. 2007, \mnras, 374, 910
\bibitem[Fan et al.(2008)]{fan06} Fan, X. et al. 2006, \aj, 132, 117
\bibitem[Ferrara et al.(1991)]{ferr91} Ferrara, A., Ferrini, F., Barsella, B. \& Franco, J. 1991, \apj, 381, 137
\bibitem[Finkelstein et al.(2007)]{f07} Finkelstein, S. L., Rhoads, J. E., Malhotra, S., Pirzkal, N. \& Wang, J. X. 2007, \apj, 660, 1023
\bibitem[Finkelstein et al.(2008)]{f08} Finkelstein, S. L., Rhoads, J. E., Malhotra, S., Grogin, N. \& Wang, J. X. 2008, \apj, 678, 655
\bibitem[Finkelstein et al.(2009)]{f09} Finkelstein, S. L., Rhoads, J. E., Malhotra, S. \& Grogin, N. 2009, \apj, 691, 465
\bibitem[Finkelstein et al.(2009b)]{f09b} Finkelstein, S. L. et al. 2009, \apj, 700, 376
\bibitem[Fosbury et al.(2003)]{fos03} Fosbury, R. A. E., et al. 2003, \apj, 596, 797
\bibitem[Fukugita et al.(2010)]{fuk10} Fukugita, M., Yasuda, M., Doi, M., Gunn, J. E. \& York, D. G. 2010, submitted to AJ
\bibitem[Gawiser et al.(2006)]{gaw06} Gawiser, E. et al. 2006, \apj, 642, L13
\bibitem[Gawiser et al.(2007)]{gaw07} Gawiser, E. et al. 2007, \apj, 671, 278
\bibitem[Hansen \& Oh(2006)]{ho06} Hansen, M. \& Oh, S. P. 2006, MNRAS, 367, 979
\bibitem[Hayes et al.(2010)]{ha10}Hayes, M. et al. 2010, Nature, 464, 562
\bibitem[Heckman et al.(1990)]{he90} Heckman, T. M., Armus, L., Miley, G., K. 1990, \apjs, 74, 833
\bibitem[Heckman et al.(2000)]{he00} Heckman, T. M., Lehnert, M., D., Strickland, D., K., M., Armus, L. 2000, \apjs, 129, 493
\bibitem[Heckman et al.(2001)]{he01} Heckman, T. M., Semback, K., Meurer, G., Leitherer, C., Calzetti, D. \& Martin, C. 2001, \apj, 558, 56
\bibitem[Heckman(2002)]{he02} Heckman, T. M. 2002, ASPCS, 254, 292
\bibitem[Holden et al.(2001)]{ho01} Holden, B. P., et al. 2001, AJ, 122, 629 
\bibitem[Holwerda(2005)]{ho05} Holwerda, B. W. 2005, astro-ph/0512139 
\bibitem[Ilbert, Capak \& Salvato et al.(2009)]{ics09} Ilbert, O., Capak P., \& Salvato, M. et al. 2009, \apj, 690, 1236 
\bibitem[Kornei et al.(2010)]{korn10} Kornei, K. A., et al. 2010, \apj, 711, 693
\bibitem[Kashikawa et al.(2006)]{kash06} Kashikawa, N. et al. 2006, \apj, 648, 7
\bibitem[Lai et al.(2008)]{lai08} Lai, K. et al. 2008, ApJ, 674, 70
\bibitem[Madau(1995)]{mad} Madau, P. 1995, ApJ, 441, 18
\bibitem[Malhotra \& Rhoads(2006)]{mr06} Malhotra, S. \& Rhoads, J. E. 2006, ApJ, 647, L95
\bibitem[Malhotra \& Rhoads(2004)]{mr04} Malhotra, S. \& Rhoads, J. E. 2004, ApJ, 617, L5
\bibitem[Malhotra et al.(2010)]{ma10} Malhotra, S. et al. 2010 (in prep)
\bibitem[McCandliss(2009)]{mcc09} McCandliss, S. R. 2009, AIPCS, 1135, 309
\bibitem[McDonald et al.(2000)]{mcd00}McDonald, P., Miralda-Escud\'{e}, J., Rauch, M., Sargent, W. L. W.,  Barlow, T. A., Cen, R., \& Ostriker, J. P. 2000, \apj\ 543, 1
\bibitem[Miyazaki et al.(2002)]{mi02} Miyazaki, S. et al. 2002, \pasj, 54, 833
\bibitem[Neufeld(1991)]{neu91} Neufeld, D. A. 1991, ApJ, 370L, 85
\bibitem[Nilsson et al.(2007)]{nil} Nilsson, K. K. et al. 2007, A\&A, 471, 71
\bibitem[Ono et al.(2010)]{ono10} Ono, Y. et al. 2010, \apj, 724, 1524
\bibitem[Ouichi et al.(2008)]{ou08} Ouichi, M. et al. 2008, \apjs, 176, 301
\bibitem[Partridge \& Peebles(1967)]{pp67} Partridge, R. B. \& Peebles, P. J. E. 1967, \apj, 147, 868
\bibitem[Pettini et al.(2001)]{pett01} Pettini, M., Shapley, A. E., Steidel, C. C., Cuby, J.-G., Dickinson, M., Moorwood, A.F.M., Adelberger, K. L., Giavalisco, M. 2001, \apj, 554, 981
\bibitem[Pickles(1998)]{pic98} Pickles, A. J. 1998, \pasp, 110, 863
\bibitem[Pirzkal et al.(2007)]{pir07} Pirzkal, N. et al. 2007, ApJ, 667, 49
\bibitem[Reddy et al.(2008)]{re08} Reddy, N. A. et al. 2008, ApJS, 175, 48
\bibitem[Rhoads et al.(2000)]{rho00} Rhoads, J. E., Malhotra, S., Dey, A., Stern, D., Spinrad, H. \& Jannuzi, B. T. 2000, \apj, 545, L85
\bibitem[Rhoads \& Malhotra(2001)]{rho01} Rhoads, J. E. \& Malhotra, S. 2001, \apj\, 563, L5
\bibitem[Rhoads et al.(2003)]{rho03} Rhoads, J. E., Dey, A., Malhotra, S.,  Stern, D., Spinrad, H. \& Jannuzi, B. T. Dawson, S., Brown, Michael, J. I., Landes, E. 2003, \aj, 125, 1006
\bibitem[Rhoads et al.(2009)]{rho09} Rhoads, J. E. et al. 2009, \apj, 697, 942
\bibitem[Rousselot et al.(2000)]{rou00} Rousselot, P., Lidman, C., Cuby, J. G., Moreels, G., \& Monnet, G. 2000, A\&A, 354, 1134
\bibitem[Santos(2004)]{san04} Santos, M.,R. 2004, MNRAS, 349, 1137
\bibitem[Schaerer \& de Barros(2009)]{sch09} Schaerer, D. \& de Barros, S. 2009, A\&A, 502, 423
\bibitem[Schaerer \& Verhamme(2008)]{sv08} Schaerer, D. \& Verhamme, A. 2008, A\&A, 480, 369
\bibitem[Seifert et al.(2002)]{sei02} Seifert, W. et al. 2002, ``LUCIFER: a Multi-Mode NIR Instrument for LBT'',  SPIE Proc. Vol. 4841 Instrument Design and Performance for Optical/Infrared Ground-Based Telescopes,
Moorwood, Iye (Eds.), 962
\bibitem[Shapley et al.(2001)]{shap} Shapley, A. E., Steidel, C. C., Adelberger, K. L., Dickinson, M., Giavalisco, M. \&  Pettini, M. 2001, \apj, 562, 95
\bibitem[Shapley et al.(2003)]{shap03} Shapley, A. E., Steidel, C. C., Pettini, M. \& Adelberger, K. L. 2003, \apj, 588, 65
\bibitem[Shimasaku et al.(2006)]{shima06} Shimasaku, K. et al. 2006, \pasj,58, 313
\bibitem[Stark et al.(2010)]{star10} Stark, D. P., Ellis, R. S., Chiu, K., Ouchi, M. \& Bunker, A. 2010, MNRAS, 408, 1628
\bibitem[Somerville \& Primack(1999)]{sp99} Somerville, R. S. \& Primack, J. R. 1999, \mnras, 310, 1087
\bibitem[Spergel et al.(2007)]{sper} Spergel, D. N. et al. 2007, ApJS, 170, 377
\bibitem[Steidel et al.(2000)]{stei00} Steidel, C. C., Adelberger, K. L., Shapley, A. E., Pettini, M., Dickinson, M., Giavalisco, M. 2000, \apj, 532, 170
\bibitem[Steidel et al.(2001)]{stei01} Steidel, C. C., Pettini, M., Adelberger, K. L. 2001, \apj, 546, 665
\bibitem[Steidel et al.(2010)]{stei10} Steidel, C. C. et al. 2010, \apj, accepted
\bibitem[Stern et al.(2005)]{s05} Stern et al. 2005, \apj, 619, 12
\bibitem[Tapken et al.(2007)]{tap07} Tapken, C., Appenzeller, I., Noll, S., Richling, S., Heidt, J., Meinköhn, E. \& Mehlert, D. 2007, \aa, 467, 63
\bibitem[Tenorio-Tagle et al.(1999)]{tt99} Tenorio-Tagle, G., Silich, S. A., Kunth, D., Terlevich, E., Terlevich, R. 1999, \mnras, 309, 332
\bibitem[Valdes(1993)]{val93} Valdes, F., Guide to the Kitt Peak Coude Slit Reduction Task DOSLIT, Central Computer Services, NOAO, 1993
\bibitem[Venemans et al.(2005)]{ven05} Venemans, B. P. et al. 2005, A\&A, 431, 793
\bibitem[Verhamme et al.(2006)]{v06} Verhamme, A., Schaerer, D. \& Maselli, A. 2006, A\&A, 460, 397
\bibitem[Verhamme et al.(2008)]{v08} Verhamme, A., Schaerer, D., Atek, H., \& Tapken, C. 2008, A\&A, 491, 89
\bibitem[Williams et al.(2004)]{will04} Williams, G. G., Olszewski, E. \& Lesser, M. P. 2004, SPIE, 5492, 787
\bibitem[Wright(2006)]{wri06} Wright, E. L. 2006, PASP, 118, 1711
\bibitem[Zheng et al.(2010)]{zhe02} Zheng, Z., Cen, R., Hy, T., \& Miralda-Escude, J. 2010, \apj, 715, 574

\end{thebibliography}
\end{document}